\documentstyle[twoside,fleqn,espcrc2,epsfig]{article}
\newcommand{\bm}[1]{\mbox{\boldmath $#1$}}

\title{Vector decay constants in quarkonia}

\author{B.D.~Jones\thanks{Presented by B.D.~Jones
at Lattice '98, 13--18 July.}
         and R.M.~Woloshyn \\\vspace{2ex}
        TRIUMF, 
        4004 Wesbrook Mall, 
        Vancouver, British Columbia, Canada, V6T 2A3} 
        
\begin{document}

\begin{abstract}
Lattice NRQCD with leading finite lattice spacing errors removed is used to simulate
 heavy-heavy vector decay constants. Quenched simulations are performed at three values
of the coupling and fifteen values of the quark mass.
The improved gauge action with plaquettes and 
rectangles is used. Landau link improvement is used throughout.
``Perturbative" and nonperturbative meson masses are compared. One-loop
perturbative matching between lattice and continuum heavy-heavy vector currents is performed.
The data is consistent with $a f_V \propto \sqrt{M_V\, a}$.
\end{abstract}

\thispagestyle{myheadings}
\markright{TRI-PP-98-24}

\maketitle

\section{INTRODUCTION}

 The gross features of
heavy quarkonia are well described by lattice NRQCD [1--3]. 
However, spin splittings (which are
small for quarkonia) tend to be underestimated---even when relativistic corrections are 
included \cite{howard}. Spin splittings are one measure of the mesonic 
wave function at the
origin;
another is the vector meson
decay constant.

A previous study \cite{davthack} of the vector decay constant found large corrections 
from the perturbative matching between the
continuum and lattice matrix elements. Another study \cite{bodwin}
reported a rather imprecise value
for simulations with the order $v^2$ classically improved vector current.
In this paper, 
we try to improve these calculations by removing the leading finite 
lattice spacing 
errors in the fermion and gauge actions in a symmetric
 fashion and by performing a more precise simulation with the inclusion of the 
 order $v^2$ classically improved current.

\vspace{-1ex}
\subsection{Lattice NRQCD}

The fermion Lagrangian is discretized in a symmetric fashion with the leading finite lattice spacing
errors in the spatial and temporal derivatives removed---all links are tadpole improved by dividing by
$u_0$, the average link in the Landau gauge:

\vspace{-3.5ex}
\begin{eqnarray}
a {\mathcal{L}}_F &=&\psi_t^\dagger \psi_t - 
 \psi_t^\dagger \left(1-\frac{a \delta H}{2}\right)_t
\left(1-\frac{a  H_0}{2 n}\right)_t^n \nonumber \\
&&\!\!\!\!\!\!\!\!\!\!\!\!\!\!\!\!\!\!\!\!\!\!\!\!\!\!\!\!\times \frac{U_{4,t-1}^\dag }{u_0}
\left(1-\frac{a  H_0}{2 n}\right)_{t-1}^n \left(1-\frac{a \delta H}{2}\right)_{t-1} 
\psi_{t-1}\,,\label{lf}\end{eqnarray}

\begin{equation}
H_0 = -\frac{\Delta^{(2)}}{2 m}\;,\;\delta H = \frac{a^2 \Delta^{(4)}}{24 m} - 
\frac{a \left ( \Delta^{(2)}
\right)^2}{16 n m^2}
\,.
\end{equation}

\vspace{-1ex}
\noindent
$n$ is the stability parameter  chosen to satisfy $n > \frac{3}{m a}$. $\Delta^{(2)}$ is the
gauge-covariant lattice Laplacian, and $\Delta^{(4)}$ is the gauge-covariant lattice quartic
operator ($\sum_i D_i^4$).

The  gauge action is tadpole improved with
leading finite lattice spacing errors removed by rectangles:

\vspace{-4ex}
\begin{eqnarray}
S_G &=& \beta \sum_{pl} \frac{1}{3} Re \,Tr \left(1-U_{pl}\right)\nonumber \\
&&\hspace{8mm}\mbox{}-\frac{\beta}{20 u_0^2}
\sum_{rt} \frac{1}{3} Re \,Tr \left(1-U_{rt}\right) \,.
\label{glue}
\end{eqnarray}

\vspace{-1ex}
The `kinetic' meson mass is defined by

\vspace{-2ex}
\begin{equation}
E(\bm{p})-E(\bm{0})=\frac{{\bm{p}}^2}{2 M_{kin}}\,,
\end{equation}
\vspace{-1ex}
where $E(\bm{p})$ is the simulated meson energy.

\vspace{-1ex}
\subsection{Perturbative lattice NRQCD}

\begin{figure}[htb]
\epsfig{figure=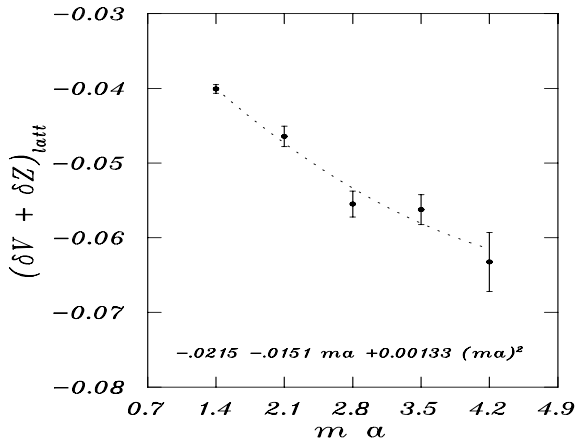,height=2.25in,width=75mm}
\vspace{-8ex}
\caption{Perturbative lattice NRQCD one-loop 
contribution to the  matching of the heavy-heavy vector current.
These
results must be multiplied by $g^2$.}
\label{fig1}
\vspace{-4ex}
\end{figure}

\begin{table*}[hbt]
\setlength{\tabcolsep}{1.3pc}
\newlength{\digitwidth} \settowidth{\digitwidth}{\rm 0}
\catcode`?=\active \def?{\kern\digitwidth}
\caption{$M_{pert}$ versus $M_{kin}$ for the runs nearest the charm and bottom regions respectively.}
\label{tab:1}
\begin{tabular*}{\textwidth}{cclllll}
\hline
           &&      & 
                 & \multicolumn{2}{l}{$M_{pert} a$}& \\
 \cline{5-6}
   \multicolumn{1}{c}{$\beta$}              
    & \multicolumn{1}{c}{$ a$(fm)} &
     \multicolumn{1}{l}{$m a [n]$}
                 & \multicolumn{1}{l}{$M_{kin} a$} 
                 & \multicolumn{1}{l}{w/o tad imp} 
                 & \multicolumn{1}{l}{w tad imp} &
     \multicolumn{1}{l}{$M_{kin}$(GeV)}        \\
\hline
$7.2$   &	$.240(3)$	& $1.6 [3] $ & $3.37(6) $ & $2.89(7)   $ & $3.60(7)   $ &$2.77(6)$\\
$7.3$   &  	$.205(3)$	& $1.5 [3]$ & $3.20(5) $ & $2.64(6)   $ & $3.37(6)   $ &$3.08(6)$\\
$7.4$   &    $.178(3)$	& $1.4 [3]$ & $3.00(6) $ & $2.39(6)   $ & $3.15(6)   $ &$3.32(9)$\\
$7.2$ &	$.198(2)$	& $4.4 [2]$ & $9.07(21)$ & $9.47(68)  $ & $8.98(68)  $ &$9.05(23)$\\
$7.3$   &    $.174(3)$	& $4.3 [2]$ & $8.74(20)$ & $9.17(61)  $ & $8.74(61)  $ &$9.91(28)$\\
$7.4$ &	$.151(2)$	& $3.5 [2]$ & $7.05(20)$ & $7.22(38)  $ & $7.06(38)  $ &$9.20(29)$\\
\hline
\end{tabular*}
\end{table*}

The Feynman rules are derived from the Lagrangians of Eqs.~(\ref{lf}) and (\ref{glue})
by making the replacement $U_\mu(x)\rightarrow \exp[i a g A_\mu^a(x)T^a]$. The gluon
propagator follows from the quadratic piece of the gauge action in Eq.~(\ref{glue})---we take
the piece proportional to $\delta_{\mu\,\nu}$ in this paper.

The continuum and lattice decay constants in one-loop perturbation theory are related by
$
f_V=Z_{match} f_{V,latt}
$,
where $
Z_{match}=
1+g^2 \left(
-\frac{2}{3 \pi^2}-(\delta V + \delta Z)_{latt}
\right)$.
Note that a linear infrared divergence cancels between the continuum and lattice
perturbative shifts. Fig.~\ref{fig1} shows
the results of the one-loop calculation. The shown errors are estimates of the extrapolation
errors. All 
one-loop integrals are performed by VEGAS \cite{vegas} in this paper.

The ``perturbative" meson  mass is defined by
$
M_{pert}=2 (m Z_m-E_0)+E_{sim}$.
Table~\ref{tab:1} shows the simulation parameters near the physical regions and compares
$M_{pert}$ and $M_{kin}$. The lattice spacing is fixed by the S--P splitting: 458~MeV,
a boosted coupling is used:
$
g^2=\frac{5}{3} \frac{6}{\beta u_0^4}
$, and $u_0=1-\frac{g^2}{4\pi}u_0^{(2)}$ is used to define $u_0^{(2)}$.

\vspace{-1ex}
\subsection{Decay constants}

The asymptotic form of a meson propagator is

\vspace{-4ex}
\begin{eqnarray}
G(\bm{p},t)&
\begin{picture}(40,6)(0,0)
\put(0,2){\vector(1,0){40}}
\put(0,-4){\mbox{\scriptsize $t-t_0\rightarrow \infty$}}
\end{picture}
&\left|
\langle 0 | J(0) | \bm{p}\rangle
\right|^2\nonumber\\
&&\mbox{} \times\exp [-E(\bm{p})(t-t_0)]\,,
\end{eqnarray}

\vspace{-1ex}
\noindent
where $J(\bm{x},t)=\chi_x \Gamma_x \psi_x$ is a non-relativistic  current
with $\Gamma_x=\Omega_x \gamma_x$, where $\Omega_x$ interpolates the meson of interest 
and $\gamma_x$ is a smearing
operator chosen in a gauge-invariant fashion:
$
\gamma_x~=~[1+\epsilon \Delta^{(2)}(x)]^{n_s}\,.
$
We set $\epsilon=1/12$ and tune the  smearing parameter $n_s$ to maximize the
overlap with the state of interest. We find the range 7--30 for $n_s$ to be sufficient, with
the P-wave requiring about twice as much smearing as the S-wave 
 and the smearing parameter increasing for decreasing
quark mass.

The above matrix element for a vector at rest is related to its  decay constant $f_V$
through
$
\langle 0 | J(0) | V\rangle=\frac{f_V M_V}{\sqrt{2 M_V}}
$,
where a non-relativistic norm is used.

Finally, note that simulations are performed with an order $v^2$ 
classically improved
current.  The improved interpolating operator for a non-relativistic vector meson is given by

\vspace{-4ex}
\begin{eqnarray}
\Omega_V^{imp}&=&\sigma_i+\frac{1}{8 m^2} \left(
\Delta^{(2)\dagger}\sigma_i+\sigma_i\Delta^{(2)}
\right)\nonumber\\
&&\mbox{}-\frac{1}{4m^2}\left(\bm{\sigma}\cdot{\bm{\Delta}}^\dagger\right)\sigma_i
\left(\bm{\sigma}\cdot\bm{\Delta}\raisebox{0ex}[2ex][1ex]{}\right)\,,
\label{imp}
\end{eqnarray}

\vspace{-1ex}
\noindent
where $\bm{\Delta}$ is the symmetric lattice gauge-covariant derivative.

\section{RESULTS AND DISCUSSION}

The data sample includes 1600 quenched gauge field configurations at $\beta =$ 7.2 and 
7.3 ($8^3\times14$), and 1200
configurations at $\beta =$ 7.4 ($10^3\times 16$). 
Multiple sources along the spatial diagonal are used to measure the
 local-smeared meson correlators with the number of
smearing steps optimized to have the best overlap
with the state of interest.
All plots include bootstrap errors using twice as many bootstrap ensembles as
there are configurations.

\begin{figure}[htb]
\epsfig{figure=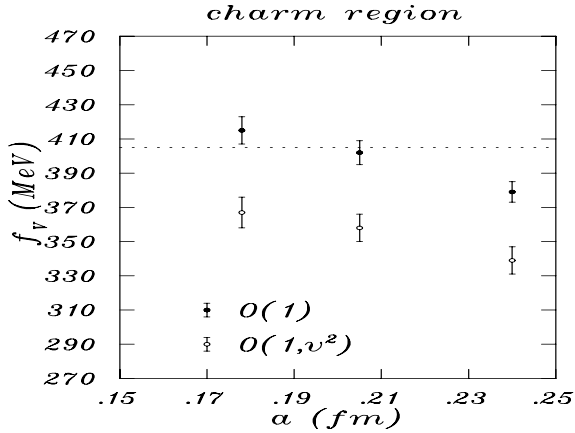,height=2.215in,width=75mm}
\vspace{-8ex}
\caption{Scaling behavior of the charm 
decay constants including
the perturbative matching. The dotted line
is
the experimental result.}
\label{fig2}
\vspace{-4ex}
\end{figure}
\begin{figure}[htb]
\epsfig{figure=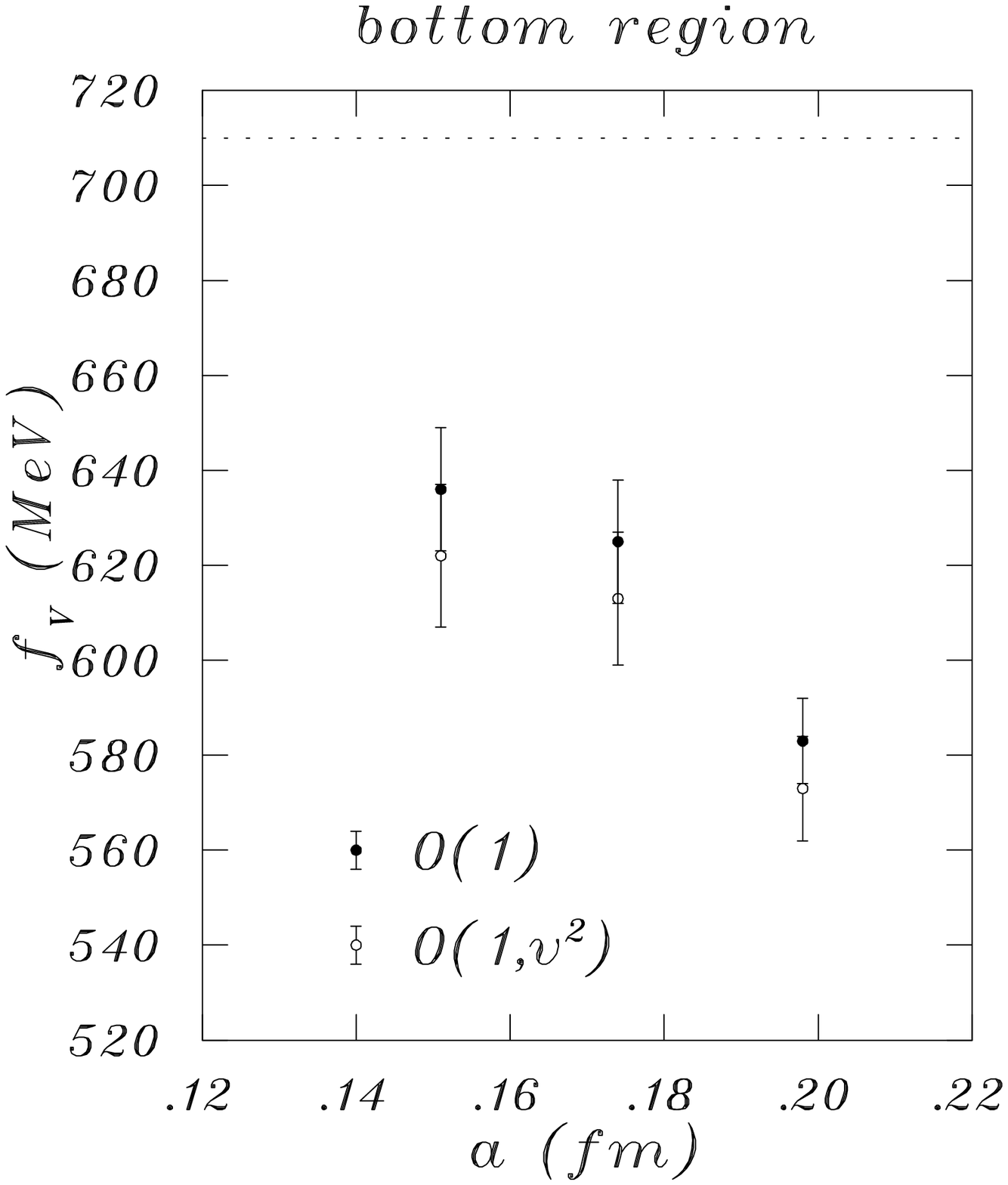,height=2.215in,width=75mm}
\vspace{-8ex}
\caption{Scaling behavior of the bottom
decay constants including
the perturbative matching. The dotted line
is
the experimental result.}
\label{fig3}
\vspace{-4ex}
\end{figure}
\begin{figure}[htb]
\vspace{2ex}
\epsfig{figure=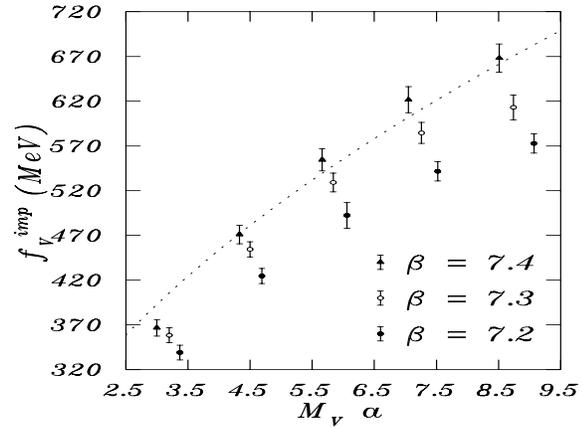,height=2.2in,width=75mm}
\vspace{-8ex}
\caption{Meson mass dependence of the order $v^2$ improved decay constant including
the perturbative matching. The dotted line is proportional to $\sqrt{M_V \,a}$.}
\label{fig4}
\vspace{-5ex}
\end{figure}

Figs.~\ref{fig2}--\ref{fig4} show our main results.
Note the 10\% underestimations (order $v^2$ improved data) as expected from the 
previous work on the
spin splittings.
 Also note  the approximate $\sqrt{M_V\,a}$ dependence. Shortly after the 
 discovery of charm, Yennie \cite{sakurai} noticed 
 this same dependence from empirical data:
\vspace{-1ex}
\begin{equation}
\frac{\Gamma^V_{e{\overline e}}}{e_q^2}\sim \mbox{constant} \sim 12 \;\mbox{keV}
\end{equation}
\vspace{-1ex}
for light through heavy ground-state vector mesons ($e_q$ is the quark charge in 
units of $e$). This is the same as our data 
since the leptonic width is proportional to $
e_q^2\frac{f_V^2}{M_V}
$.
 A final note is that 
 a \emph{linear}
 (\emph{Coulomb}) potential  implies a 
\emph{constant} (\emph{linear})
 dependence on the meson mass for the decay constant, so our data is consistent with
 a superposition of the two potentials.
 
\vspace{2ex}
\noindent
\textbf{ACKNOWLEDGMENTS}
\vspace{2ex}

The authors thank N. Shakespeare and H. Trottier for discussions. This work is supported
by the Natural Sciences and Engineering Research Council of Canada.

\thispagestyle{myheadings}
\markright{}

\end{document}